\documentstyle[multicol,epsfig,prc,aps]{revtex} 
\def\smh{\hspace*{0.3cm}}

\begin{document}

\draft    


\title{ 
Hadron Spectra and
QGP Hadronization in Au+Au Collisions at RHIC
} 

\author{
{\bf K.A. Bugaev}$^{a,b}$,
{\bf M. Ga\'zdzicki}$^c$ and
{\bf M.I. Gorenstein}$^{a,d}$ 
}

\address{ 
$^a$ Bogolyubov Institute
for Theoretical Physics,
Kiev, Ukraine\\
$^b$ Gesellschaft f\"ur Schwerionenforschung (GSI), Darmstadt, Germany\\
$^c$ Institut f\"ur  Kernphysik, Universit\"at  Frankfurt,
Germany\\
$^d$ Institut f\"ur Theoretische Physik, Universit\"at  Frankfurt,
Germany
}

\date{\today}
 
\maketitle

\begin{abstract}
\noindent
The transverse mass spectra of 
$\Omega$ hyperons and $\phi$ mesons measured recently
by STAR Collaboration in Au+Au
collisions 
at $\sqrt{s_{NN}} = 130$~GeV
are described within a hydrodynamic model 
of the quark gluon plasma 
expansion
and hadronization. The flow parameters
at the plasma  
hadronization extracted by fitting these data  are used  to
predict the transverse mass spectra 
of $J/\psi$ and $\psi^{\prime}$ mesons.
\end{abstract}

\vspace{0.2cm}

\begin{multicols}{2}

\vspace*{-0.5cm}

Recent measurements \cite{na49} of the energy dependence of pion and kaon
production in central collisions of heavy nuclei (A+A) indicate 
\cite{GaGo0}
that the transient state of deconfined matter is created
at the early stage of these collisions for energies
higher than about 40 A$\cdot$GeV, i.e. at high SPS and
RHIC energies.
Analysis of the hadron multiplicities measured in 
these collisions 
within statistical models
at the SPS \cite{HG} and RHIC \cite{HG1} energies
shows that
the chemical freeze--out  takes place
near the boundary between the quark--gluon and hadron phases.
The values of the
temperature parameter extracted from the data are similar
for both energies, $T_H=170\pm 10$~MeV, and they are 
close to the value of the deconfinement transition temperature  
at zero baryonic density estimated 
in the lattice QCD (see e.g. Ref.~\cite{karsch}).

The analysis of the experimental  data at the SPS 
\cite{Ka,Wi:99} 
and a numerical modeling of the hadron cascade stage in A+A collisions
at SPS and RHIC energies
\cite{BD,Sh} 
indicate that the kinetic (i.e., particle spectra) freeze--out
of the most abundant hadrons
takes place at  temperatures significantly lower than $T_H$.
Nevertheless,
one  expects that the kinetic freeze--out
of some heavy and weakly interacting  hadrons 
(e.g. $\Omega$ hyperons and 
$\phi$, $J/\psi$, $\psi^{\prime}$ mesons) 
may occur directly at the
quark--gluon plasma (QGP) hadronization stage or close to it.
Thus
for these hadrons the chemical and kinetic freeze--outs
coincide and are determined by the features of the QGP
hadronization.
For $\Omega$ hyperons and $\phi$ mesons this expectation is based on the
results of ``hydro QGP + hadron cascade'' approach \cite{BD,Sh}. For
$J/\psi$ and  $\psi^{\prime}$ mesons this is our suggestion
\cite{BGG,Go:02,BUG:02},
which is a straightforward consequence of the 
recently proposed
statistical mechanism of 
charmonia production at the QGP hadronization 
\cite{gago,br,go,rapp}.
Within this approach the transverse mass 
spectra of the $\Omega$,
$J/\psi$ and $\psi^{\prime}$ measured 
in Pb+Pb collisions at the SPS have been
described successfully \cite{Go:02}. 
The transverse mass spectra of $\Omega$ hyperons \cite{QM02}
and $\phi$ mesons \cite{phi} produced 
in central Au+Au collisions at $\sqrt{s_{NN}} = 130$~GeV 
were recently measured by the STAR Collaboration.
These data allow to test our model in the new
energy regime.
They also give  a unique opportunity to extract
parameters of the QGP hadronization at RHIC energies 
and consequently predict
spectra of
$J/\psi$ and $\psi'$ mesons.

Within a hydrodynamical approach of the QGP hadronization
the transverse mass spectrum of $i$-th hadron in the central rapidity
region can be written as (see, e.g., Ref.~\cite{gy}): 
\begin{eqnarray}\label{hydro}
&&\frac{dN_i}{m_T dm_T dy}~\biggl|_{y = 0}~
= 
\frac{d_i~ \lambda_i~ \gamma_i^{n_i} }{\pi}~\tau_H ~ R_H^2~ \times \nonumber \\
&&\int_{0}^{1} \xi~ d\xi~ K_1 \left(
\frac{m_T \cosh y_T}{T_H}  \right) ~  
I_0\left({ \frac{p_T\sinh y_T}{T_H}}\right)~,
\end{eqnarray}
%
where $y$ is the particle longitudinal rapidity and  $y_T(\xi)=\tanh^{-1}v_T$
is the fluid transverse rapidity.
$R_H$  and $\tau_H$ are, respectively,
the transverse system size and proper time at the hadronization (i.e., 
at the boundary between the mixed phase and hadron matter), $\xi=r/R_H$
is a relative transverse coordinate.
The particle degeneracy and fugacity are denoted as $d_i$ and $\lambda_i$,
respectively,
$m_T=\sqrt{p_T^2 + m_i^2}$ is the
hadron
transverse mass, 
$ K_1$ and  $I_0$ are the modified Bessel functions.
Parameter $\gamma_i$ in Eq.~(\ref{hydro}) ($\gamma_S$ \cite{Raf} for
$i=\phi, \Omega$ and $\gamma_C$ \cite{br,go} for $i=J/\psi$,
$\psi^{\prime}$) describes a possible deviation of strange and charm hadrons
from complete chemical equilibrium ($n_i=2$ for
$\phi,J/\psi,\psi^{\prime}$ and $n_i=3$ for $\Omega$). 

The spectrum (\ref{hydro}) is obtained under
the assumption that the hydrodynamic expansion is longitudinally boost
invariant and
that
the
freeze--out occurs at constant longitudinal proper time $\tau
=\sqrt{t^2-z^2}$
($t$ is the time and $z$ is the longitudinal coordinate), i.e. the
freeze--out time $t$ is independent of the transverse coordinate $r$. 
In order to complete Eq.~(1) the functional form of the 
transverse rapidity distribution of hadronizing matter
$y_T(\xi)$ has to be given. A linear flow profile,
$y_T(\xi)=y_T^{max}\cdot \xi$, 
used in our model 
is justified by the numerical calculations of
Ref.~\cite{Sh}.

Thus, in our model,  the QGP hadronization is described  by the
following parameters:
temperature $T_H$, ``volume" $\tau_H R_H^2$,
maximum flow rapidity  $y_T^{max}$, fugacities
$\lambda_i$,  and saturation factors $\gamma_i$.
Note that the $\phi,J/\psi,\psi^{\prime}$ have no conserved charges
and $\lambda_i=1$ for these particles.
We use the fixed values of the parameters
$T_H = 170$~MeV, $\gamma_S = 1.0$, $\lambda_{\Omega^-} = 1/
\lambda_{\Omega^+} = 1.09$ 
(note that $\lambda_{\Omega^-}\equiv \exp[(\mu_B-3\mu_S)/T]$,
where $\mu_B$ and $\mu_S$ are, respectively, baryon and
strange chemical potentials). These
(average) values of the {\it chemical
freeze-out} parameters have been found in the 
hadron gas analysis \cite{HG1} of the full set of 
the midrapidity particle number ratios measured in 
central Au+Au collisions at $\sqrt{s_{NN}} = 130$~GeV.
The fit to the  $m_T$-spectra of $\Omega^{\pm}$ hyperons \cite{QM02}
and $\phi$ mesons \cite{phi}  measured in
central (14\% for $\Omega^{\pm}$ and
11\% for $\phi$) Au+Au collisions at $\sqrt{s_{NN}} = 130$~GeV 
is shown in Fig.~1.
The fit results are: 
$y_T^{max} = 0.74 \pm 0.09$, 
$\tau_H R_H^2 =275 \pm 70 $ fm$^3$/c
and
$\chi^2/ndf \cong $ 0.46. 
In the calculation of errors of the two free parameters of the model 
the uncertainties of $T_H$ ($\pm 5$ MeV),
$\gamma_S$ ($\pm 0.05$) and $\lambda_{\Omega^-}$ ($\pm 0.06$)
were taken into account.

\vspace*{0.2cm}

\begin{figure}
\mbox{ \epsfig{file=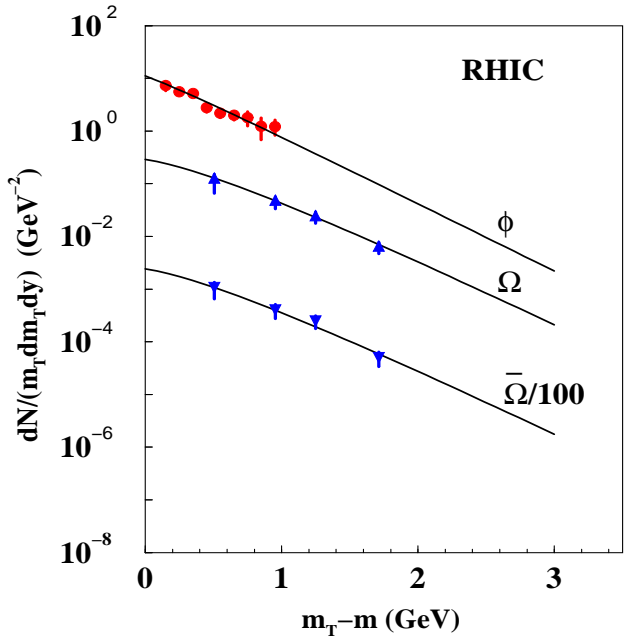,width=80mm,height=85mm} }
\caption{
The hadron transverse mass  spectra
in  Au+Au collisions at $\sqrt{s_{NN}} = 130$~GeV are shown.
The points indicate experimental data
for the $\Omega$
\protect\cite{QM02}
and
$\phi$
\protect\cite{phi}
measured by STAR.
The model results 
are shown by full lines.
}
\label{fig1}
\end{figure}

\vspace*{-0.2cm} 

A simple exponential approximation 
of the spectra
is usually utilized to parameterize the experimental data: 
\vspace*{-0.2cm} 

\begin{equation}\label{T*1}
\frac{dN}{m_T dm_T dy}~\biggl|_{y = 0}~
= ~
C~
\exp\left( -~\frac{m_T}{T^*} \right)~.
\end{equation}

\vspace*{-0.3cm}

\noindent
Note that in
Refs.\cite{BGG,Go:02,BUG:02}
an additional factor $m_T^{1/2}$ 
was present in the r.h.s. of Eq.~(\ref{T*1}).
It led  to smaller 
values of $T^*$ when fitting the same spectrum.
The $m_T$--spectrum (1) may, however, deviate significantly 
from a purely exponential one
and its shape depends on the
magnitude of the transverse flow and the mass of the particle.
The normalization factors $C$
and the inverse slope parameters $T^*$ in different intervals
of $m_T-m$ can be
found from the $\phi$,
$\Omega$, $J/\psi$ and $\psi^{\prime}$ spectra 
given by Eq.~(1) using the maximum likelihood method.
The average values of $T^*$ for the $m_T$ domains
of ``low-$p_T$'' ($m_T-m < 0.6$~GeV) and ``high-$p_T$'' 
($0.6~$GeV$ < m_T-m < 1.6$~GeV), 
discussed in Refs.~\cite{Sh,BUG:02},
are shown in Fig.~2. 
The values of $T^*$  obtained by fitting the $\Omega^{\pm}$, $J/\psi$ and $\psi^{\prime}$
data in Pb+Pb collisions at
158 A$\cdot$GeV 
(see Ref.~\cite{Go:02})
are also shown 
for comparison.
The observed increase of $T^*$ with
increase of the hadron mass
is much stronger at RHIC than at SPS energies.
It is caused by  larger transverse flow velocity of
hadronizing QGP  at RHIC ($\overline{v}_T\cong 0.44$) than at
SPS ($\overline{v}_T\cong 0.19$).
The increase of $T^*$ is much more pronounced in ``low-$p_T$'' region
than in ``high-$p_T$'' one.
In our model the $m_T$-spectra of charmonia
are extraordinary affected by the stronger
transverse flow at RHIC due to
enormous masses of these hadrons.
Thus, the data on $J/\psi$ and $\psi^{\prime}$ production
in Au+Au collisions, soon to be obtained at RHIC, 
should allow to test the hypothesis of their 
formation at the QGP hadronization.


\vspace*{-0.35cm}

\begin{figure}
\mbox{ \epsfig{figure=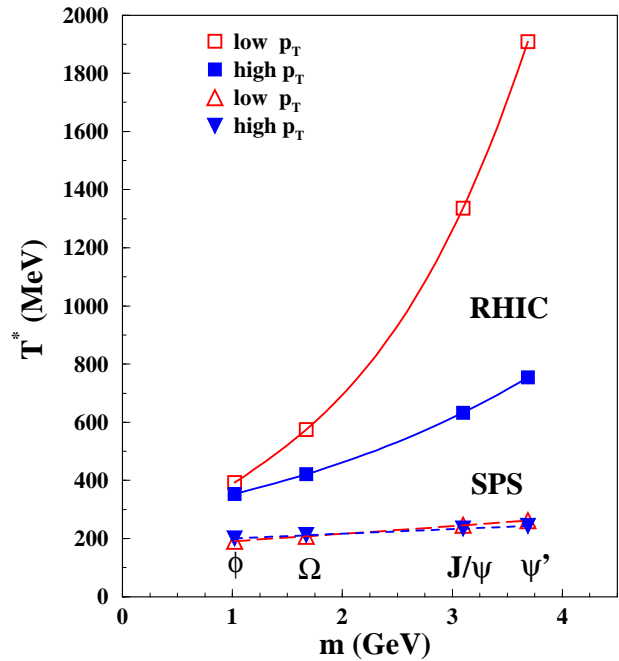,width=80mm,height=90mm} }
\caption{
The 
values of the inverse slope parameters $T^*$ for two different
$m_T$ domains -- ``low-$p_T$'' ($m_T-m < 0.6$~GeV)
and  ``high-$p_T$'' ($0.6~$GeV$ < m_T-m < 1.6$~GeV)  -- in
Au+Au collisions at $\sqrt{s_{NN}}=130$~GeV
are presented. They are found using  Eq.~(1) with $T_H=170$ MeV and
$y^{max}_T=0.74$.
For comparison, the values of $T^*$
extracted from fitting the data in Pb+Pb collisions at the
SPS (Eq.~(1) with $T_H=170$~MeV, $y_T^{max}=0.28$, see
Ref.~\protect\cite{Go:02})
are also shown.}
\label{fig2}
\end{figure}

We note here that at present
there exists an uncertainty in the estimates of the $\gamma_C$ factor,
therefore,
the predictions concerning charmonia multiplicities
in Au+Au collisions at RHIC within
statistical approaches significantly vary
and their discussion goes beyond the scope of this letter.

The ``volume parameter'' $\tau_H R^2_H \equiv A(T_H)$
extracted from the fit to the $\Omega$ and $\phi$ 
spectra defines the line $\tau_H=A(T_H)\cdot R_H^{-2}$ in the 
$R_H$--$\tau_H$ plane.
The allowed region in 
the $R_H$--$\tau_H$ plane can be estimated by varying 
the temperature parameter within its limits, 
$T_H=165$~MeV and $T_H=175$~MeV.
The resulting lines are shown in Fig.~3. 
The transverse radius $R_H = 5\div 7 $~fm and
the proper time $\tau_H = 8\div 11$~fm/c at the QGP hadronization
can be estimated from 
the hydrodynamical  calculations of \cite{Sh} for central Au+Au collisions at
$\sqrt{s_{NN}}=130$~GeV 
(see Fig.~3 in Ref.~\cite{Sh}). 
These model boundaries and their intersection with the $R_H$--$\tau_H$
region found in our  analysis are shown in
Fig. 3.


\vspace*{-0.3cm}

\begin{figure}
\mbox{ \epsfig{figure=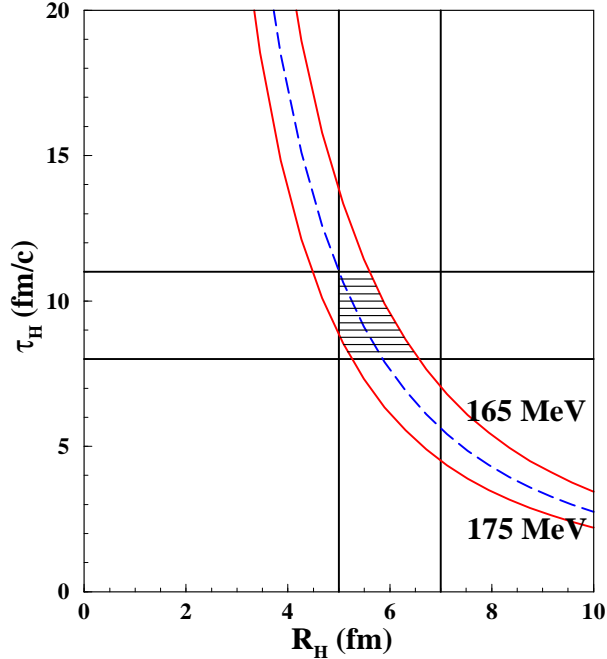,width=80mm,height=90mm} }
\caption{
The lines $\tau_H=A(T_H)\cdot R^{-2}_H$ of constant ``volume parameter''
$A(T_H)$
are shown: $T_H=170$~MeV corresponds to the dashed line, $T_H=165$~MeV and
$T_H=175$~MeV correspond to
the lower and upper solid lines, respectively.
The dashed area  is the intersection of the $R_H$--$\tau_H$
region between the $T_H=165$~MeV and $T_H=175$~MeV lines 
with the region of $R_H=5\div7$~fm and
$\tau_H=8\div11$~fm/c estimated from Ref.~\protect\cite{Sh}.
 }
 \label{fig3}
 \end{figure}

\vspace*{-0.5cm}

\begin{center}
\begin{tabular}{|c|c|c|c|}\hline

 &  $\smh T_{low-p_T}^*$ \smh  & \smh  $T_{high-p_T}^*$ \smh & \smh Refs. \smh \\
 & (MeV) & (MeV) & \\
\hline
\hline
{\bf DATA}~~~$p,\bar{p}$  & $455\pm 105 $  & $290 \pm 40$ & \cite{star1,phenix1}\\
%
%
\hline
Hydro+RQMD  & 480 & 300 & \cite{Sh} \\
\hline
Single freeze-out  & 315  & 310 & \cite{BF1,BF2} \\
\hline
\hline
{\bf DATA}~~~{\footnotesize $\Lambda,\bar{\Lambda}$} & $505\pm 60 $  & $320\pm 30$ & \cite{star2,phenix2}\\
%
%
\hline
Hydro+RQMD & 440 & 310  & \cite{Sh} \\
\hline
Single freeze-out  & 360 & 330 & \cite{BF2}
\\
\hline
\hline
\end{tabular}
\\
%

\end{center}

\noindent
 Table I. 
{\small 
The values of inverse slope parameters $T^*$ for
(anti)protons and (anti)lambdas in Au+Au collisions at
$\sqrt{ s_{NN} }=130$~GeV are presented. The experimental
values are taken as the average ones over the STAR and PHENIX results
(a difference in the results for
particle and its
anti-particle is small).
} 

Within our approach the $m_T$-spectra of
$\phi$, $\Omega$, $J/\psi,\psi^{\prime}$
are assumed to be frozen at the space-time hyper-surface
where the hadron phase starts.
This assumption is justified by
the small hadronic cross sections and large masses of these particles
(in addition, the $m_T$-spectra  of these hadrons are  almost not
affected by the resonance feeding).
However, the~ $m_T$-spectra\, of\, many other hadrons 

\begin{figure}
\mbox{ \epsfig{figure=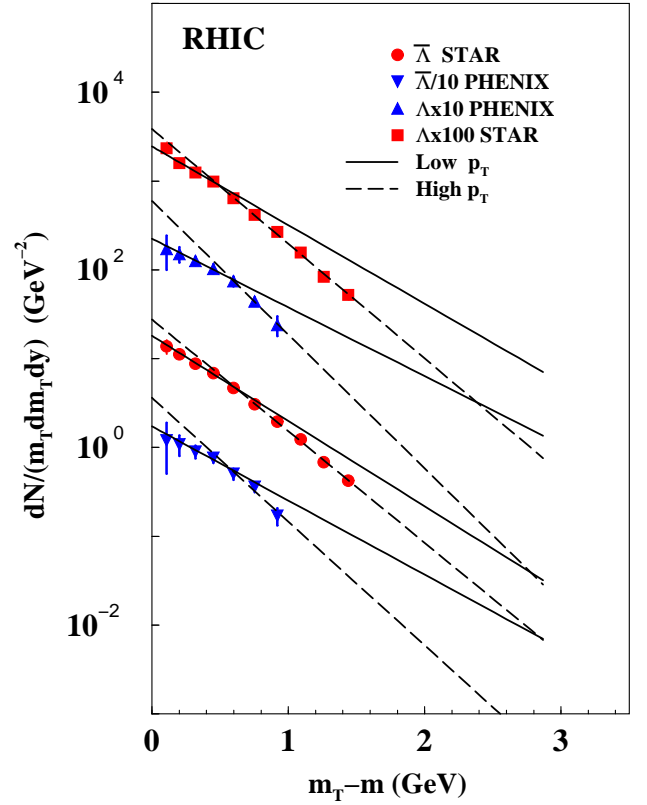,width=80mm,height=110mm} } 
\caption{
The points indicate the experimental $m_T$-spectra of
the $\Lambda$ and ${\bar \Lambda}$ in central Au+Au 
collisions  at $\sqrt{ s_{NN} }=130$~GeV measured by the STAR
\protect\cite{star2} and  PHENIX \protect\cite{phenix2}
Collaborations. The ``straight lines" are 
the exponential approximations of the spectra with 
Eq.~(\ref{T*1})
in the low-$p_T$ (solid lines) and high-$p_T$ (dashed lines) regions.}
\label{fig4}
\end{figure}

\vspace*{-0.5cm}

\noindent
are expected
to be significantly modified by hadronic rescattering.
Contrary to this expectation it was recently postulated
\cite{BF1,BF2} that  the simultaneous chemical
and kinetic freeze-out in Au+Au collisions at RHIC
occurs for all hadrons
(a single freeze-out model).
Do experimental data allow us to distinguish between
these two approaches?
 
The ``hydro QGP + hadron cascade'' approach \cite{Sh}
predicts for central  Au+Au
collisions at $\sqrt{ s_{NN} }=130$~GeV  that the hadron cascade
stage modifies the $m_T$-spectra of nucleons and
$\Lambda$ hyperons substantially. 
In particular a large increase of the inverse slope
parameter in the low-$p_T$ region is expected for these
hadrons as a result of hadronic rescattering and resonance
decay effects.
Thus measurements of (anti)proton and (anti)lambda
$m_T$-spectra should allow to distinguish between
the single freeze-out model and models which
assume different kinetic freeze-out conditions for different hadrons.
We performed the $T^*$  analysis of the present RHIC data from STAR \cite{star1,star2} and
PHENIX \cite{phenix1,phenix2}.
The resulting  $T^*$ values are
summarized in Table I together with the predictions
of the single freeze--out model \cite{BF1,BF2}
and ``QGP hydro + hadron cascade" model \cite{Sh}.
The $m_T$-spectra of the $\Lambda$ and ${\bar \Lambda}$
are also shown in Fig.~4.
There are significant systematic differences between 
$T^*$ parameters obtained from the STAR and PHENIX data.
In view of this fact, the values quoted in the Table I are
calculated as an arithmetic average of both results, whereas
the (systematic) error was estimated to be a half of the 
difference between them.

Despite 
the large uncertainties,
the data seem  to favor
the ``QGP hydro + hadron cascade" model over
the single freeze--out model.
Additional data in the low-$p_T$ region and their
theoretical analysis
would be helpful to clarify  presence of the hadron cascade stage and
its influence on $T^*_{low-p_T}$ of (anti)protons and (anti)lambdas.

The results on $m_T$--spectra of charmonia in central Au+Au
collisions at the RHIC energies
are expected to be available soon.
They should  allow to test 
a statistical approach to the charmonia production
at the QGP hadronization in high energy nuclear collisions.  
In particular, within this approach,
we predict a strong 
(a few times)
increase of the inverse slope
parameter $T^*$ of the charmonia $m_T$--spectra at RHIC
in comparison with that at SPS.
The higher is the energy the larger inverse slope is expected due to
increasing transverse flow of hadronizing QGP.
Thus, at $\sqrt{s_{NN}}=200$~GeV the increase
of $T^*$  should become even more
pronounced than at $\sqrt{s_{NN}}=130$~GeV. 
Due to strong sensitivity of
the charmonia spectra
to the hadronization temperature and transverse flow velocity,
their analysis should significantly improve our estimate of
these parameters.

\vspace*{0.4cm}

\noindent
{\bf  Acknowledgements.}
We thank A. L. Blokhin, P. Braun--Munzinger and W. Greiner
for discussions and comments.
The financial supports from the Humboldt Foundation 
and INTAS grant 00--00366 are acknowledged.


\end{multicols} 

\begin{thebibliography}{99}

\bibitem{na49}
S. V. Afanasiev {\it et al.} (NA49 Collab.),
nucl--ex/0205002, to appear in Phys. Rev. {\bf C}.

\bibitem{GaGo0}
M. Ga\'zdzicki and M.I. Gorenstein,
Acta Phys. Polon. {\bf B30}, 2705 (1999) and
references therein.


\bibitem{HG}   
J. Cleymans and H. Satz, Z. Phys. {\bf C57}, 135 (1993);
%
J. Sollfrank, M. Ga\'zdzicki, U. Heinz and J. Rafelski,
Z. Phys. {\bf C61}, 659 (1994); 
%
%
P.~Braun--Munzinger, I.~Heppe and J.~Stachel,
Phys.\ Lett.  {\bf B465}, 15 (1999);
%
%
G.~D.~Yen and M.~I.~Gorenstein, Phys.\ Rev. {\bf C59}, 2788
(1999);
%
%
F.~Becattini {\it et al.},
Phys. Rev. {\bf C64}, 024901 (2001).
%

\bibitem{HG1} 
P.~Braun--Munzinger {\it et al.},
Phys. Lett. {\bf B518}, 41 (2001);
%
%
W. Florkowski, W. Broniowski and M. Michalec,
Acta Phys. Pol. {\bf B33}, 761 (2002);
%
%
F. Becattini, J. Phys. {\bf G28}, 2041 (2002).
%

\bibitem{karsch} F. Karsch, Nucl. Phys. Proc. Suppl. {\bf 83},
14 (2000).
%


\bibitem{Ka}
B. K\"ampfer, hep--ph/9612336,
%
%
B. K\"ampfer {\it et al}.,
J. Phys. {\bf G23}, 2001 (1997).
%
 
\bibitem{Wi:99}
U.A. Wiedemenn and U. Heinz, Phys. Rep. {\bf 319}, 145
(1999) and references therein.
%


\bibitem{BD}
S. Bass and A. Dumitru, Phys. Rev. {\bf C61}, 064909 (2000).
%
 
\bibitem{Sh}
D. Teaney, J. Lauret and E.V. Shuryak, nucl--th/0110037.
%

\bibitem{BGG}
K.A.~Bugaev, M. Ga\'zdzicki and M.I. Gorenstein,
Phys. Lett. {\bf B523}, 255 (2001);
%
%
K.A.~Bugaev, J.Phys. {\bf G28} 1981 (2002). 
%

 
\bibitem{Go:02}
M.I. Gorenstein, K.A. Bugaev and  M. Ga\'zdzicki,
Phys. Rev. Lett. {\bf 88}, 132301
(2002).
%

\bibitem{BUG:02}
 K.A. Bugaev, M. Ga\'zdzicki and M.I. Gorenstein,
Phys. Lett.  {\bf B544}, 127 (2002).
%

\bibitem{gago}
M. Ga\'zdzicki and M.I. Gorenstein,
Phys. Rev. Lett. {\bf 83}, 4009 (1999).
%


\bibitem{br} P. Braun-Munzinger and J. Stachel, 
Phys. Lett. {\bf B490}, 196 (2000); Nucl. Phys. {\bf A690}, 119c (2001); 
%
%
A. Andronic, P. Braun-Munzinger, K. Redlich and J. Stachel,
nucl-th/0209035.

\bibitem{go} M.I. Gorenstein {\it et al.},  
Phys. Lett. {\bf B509}, 277 (2001); 
%
%
J. Phys. {\bf G28}, 2297 (2002) and references therein.   
%

\bibitem{rapp} L. Grandchamp and R. Rapp,
Phys. Lett. {\bf B523}, 60 (2001);
%
%
Nucl. Phys. {\bf A709}, 415 (2002);
hep-ph/0209141.
%

\bibitem{QM02}
G. van Buren [STAR Collaboration], talk given at QM2002 (2002).
 
 \bibitem{phi}
C. Adler {\it et al.}, STAR Collaboration, Phys. Rev. {\bf C65} 041901(R) (2002);
%
%
F. Laue (for the STAR Collaboration), J. Phys. {\bf G28}, 2051  (2002).
%


\bibitem{gy}
M. Gyulassy, nucl-th/0106072.
%

\bibitem{Raf}
J. Rafelski, Phys. Lett. {\bf B262}, 333 (1991).
%


\bibitem{BF1}
 W. Broniowski and W. Florkowski, Phys. Rev. Lett. {\bf 87},  272302 
(2001).
%

\bibitem{BF2}
W. Broniowski and W. Florkowski, Phys. Rev. {\bf C65}, 064905 (2002).
%


\bibitem{star1}
C. Adler {\it et al.}, STAR Collaboration, Phys. Rev. Lett. {\bf 87}, 262301 (2001).
%

%
\bibitem{star2} C. Adler {\it et al.}, STAR Collaboration, Phys. Rev. Lett.
{\bf 89}, 092301 (2002).
%



%
\bibitem{phenix1}
K.  Adcox {\it et al.}, PHENIX Collaboration, Phys. Rev Lett. {\bf 88}, 242301 
(2002).
%



%
\bibitem{phenix2}
K. Adcox {\it et al.}, PHENIX Collaboration, Phys. Rev. Lett {\bf 89}, 092302
(2002).
%




\end{thebibliography}
\end{document}